\documentclass[fleqn,usenatbib]{mnras}

\usepackage{newtxtext,newtxmath}

\usepackage[T1]{fontenc}

\DeclareRobustCommand{\VAN}[3]{#2}
\let\VANthebibliography\thebibliography
\def\thebibliography{\DeclareRobustCommand{\VAN}[3]{##3}\VANthebibliography}

\usepackage{graphicx}	
\usepackage{amsmath}	
\usepackage{gensymb}
\usepackage{subcaption}
\usepackage{bm}

\title[Faraday rotation in RMHD AGN jet simulations]{Faraday rotation as a probe of radio galaxy environment in RMHD AGN jet simulations}

\author[L. A. Jerrim et al.]{
L. A. Jerrim,$^{1}$\thanks{E-mail: larissa.jerrim@utas.edu.au}
S. S. Shabala,$^{1,2}$
P. M. Yates-Jones,$^{1,2}$
M. G. H, Krause,$^{3}$
R. J. Turner,$^{1}$
\newauthor C. S. Anderson,$^{4}$
G. S. C. Stewart,$^{1}$
C. Power,$^{2,5}$
P. E. Rodman$^{1,6}$
\\
$^{1}$School of Natural Sciences, University of Tasmania, Private Bag 37, Hobart TAS 7001, Australia \\
$^{2}$ARC Centre of Excellence for All Sky Astrophysics in 3 Dimensions (ASTRO 3D), Australia \\
$^{3}$Centre for Astrophysics Research, University of Hertfordshire, College Lane, Hatfield, Herts AL10 9AB, UK \\
$^{4}$Research School of Astronomy and Astrophysics, The Australian National University, ACT 2611, Australia \\
$^{5}$International Centre for Radio Astronomy Research, University of Western Australia, 35 Stirling Highway, Crawley, WA 6009, Australia \\
$^{6}$Institute of Astronomy, University of Cambridge, Madingley Road, Cambridge, CB3 0HA, UK \\
}

\date{Accepted XXX. Received YYY; in original form ZZZ}

\pubyear{2023}

\begin{document}
\label{firstpage}
\pagerange{\pageref{firstpage}--\pageref{lastpage}}
\maketitle

\begin{abstract}
Active galactic nuclei (AGN) play an integral role in galaxy formation and evolution by influencing galaxies and their environments through radio jet feedback. Historically, interpreting observations of radio galaxies and quantifying radio jet feedback has been challenging due to degeneracies between their physical parameters. In particular, it is well-established that different combinations of jet kinetic power and environment density can yield indistinguishable radio continuum properties, including apparent size and Stokes I luminosity. We present an approach to breaking this degeneracy by probing the line-of-sight environment with Faraday rotation. We study this effect in simulations of three-dimensional relativistic magnetohydrodynamic AGN jets in idealised environments with turbulent magnetic fields. We generate synthetic Stokes I emission and Faraday rotation measure (RM) maps, which enable us to distinguish between our simulated sources. We find enhanced RMs near the jet head and lobe edges and an RM reversal across the jet axis. We show that increasing the environment density and the average cluster magnetic field strength broadens the distribution of Faraday rotation measure values. We study the depolarisation properties of our sources, finding that the hotspot regions depolarise at lower frequencies than the lobes. We quantify the effect of depolarisation on the RM distribution, finding that the frequency at which the source is too depolarised to measure the RM distribution accurately is a probe of environmental properties. This technique offers a range of new opportunities for upcoming surveys, including probing radio galaxy environments and determining more accurate estimates of the AGN feedback budget.

\end{abstract}




\section{Introduction}
Active galactic nuclei (AGN) are the most energetic objects in the Universe and provide an important source of feedback in galaxy evolution \citep[e.g.][]{2015ARA&A..53...51S}. AGN jets are the main source of AGN feedback in the local Universe \citep{2012ARA&A..50..455F}. These jets act to heat and displace cool gas from cluster centres \citep{2007ARA&A..45..117M} and regulate star formation \citep{2011MNRAS.413.2815S,2021A&A...654A...8N}. These effects have been seen in both numerical \citep[e.g.,][]{2007MNRAS.380..877S,2013MNRAS.436.3031V} and semi-analytic \citep[e.g.,][]{2004ApJ...600..580G,2017MNRAS.471..658R,2019MNRAS.486.1509R} galaxy formation models, as well as in numerical simulations of AGN jets \citep{gaibler_jet-induced_2012,yang_how_2016,2021MNRAS.508.4738M}. 

To quantify the energetics of AGN feedback, an accurate measure of the kinetic power of radio jets is required; however, this cannot be directly inferred from the radio luminosity of the source. The size and luminosity of radio galaxies are influenced by both their intrinsic jet parameters and the ambient environment the jets interact with. Linear size—radio luminosity diagrams can be a useful tool to estimate jet power, age, and environmental properties if the relationship between these parameters is well-known \citep{1963SvA.....6..465S,1997MNRAS.292..723K,turner_energetics_2015,2019A&A...622A..12H}. \citet{2013MNRAS.430..174H,2014MNRAS.443.1482H} have shown through numerical simulations that different combinations of jet and environment parameters can result in radio galaxies with similar luminosity and linear size. Therefore, the relationship between these parameters is not simple, and another constraint is required to make inferences about kinetic jet powers.

Polarimetry is a useful tool for studying AGN jets and has been used widely to study parsec-scale jets \citep[e.g.][]{2002PASJ...54L..39A,2012AJ....144..105H,2018A&A...612A..67G} and kiloparsec-scale sources \citep[e.g.][]{2011MNRAS.413.2525G,2018MNRAS.475.4263O,2020ApJ...903...36S}. The polarisation state of synchrotron emission from radio galaxies changes as the emission travels to the observer due to Faraday rotation \citep{2021MNRAS.tmp.1703F}. The amount of Faraday rotation is quantified by the rotation measure (RM) or the Faraday depth ($\phi$) along the line of sight. For rotation by a foreground magnetized plasma, the Faraday depth is equal to the rotation measure \citep{2019SSRv..215...16V}. The polarisation angle $\chi$ is given by \citep{burn_depolarization_1966}:

\begin{equation}
    \chi = \chi_0 + \phi \lambda^2,
\end{equation}

where $\chi_0$ is the intrinsic polarisation angle, $\phi$ is the Faraday depth and $\lambda$ is the wavelength of the emission. The Faraday depth depends on the magnetic field, cluster density, and the line of sight to the source as \citep{2002ARA&A..40..319C}:

\begin{equation}
\label{eqn:RM}
    \phi = 812 \int^l_0 n_e \bm{B} \cdot \mathrm{d}\bm{l} \: \mathrm{rad\,m^{-2}},
\end{equation}

where the thermal electron number density $n_e$ is in cm$^{-3}$, the magnetic field strength $\bm{B}$ is in $\mu$G, and the path length $\mathrm{d}\bm{l}$ is in kpc. If RM $>$ 0, then on average, the magnetic field is directed towards the observer, and similarly, if RM $<$ 0, the magnetic field is directed away from the observer \citep[as defined by][]{1972ApJ...172...43M}. Generally, there is a Galactic foreground RM contribution on the order of tens to a few hundred rad/m$^2$ \citep{2022A&A...657A..43H}.

If the foreground components of the RM can be removed, the residual RM of a radio galaxy can provide insights into the physics of the system. For example, \citet{2018MNRAS.475.4263O} presented RM observations of the radio galaxy PKS J0636-2036 and concluded that the dominant contributor to the RM of this radio galaxy is external Faraday depolarisation due to either a magnetized intergalactic medium (IGM) or shock-enhanced IGM gas. Previously, \citet{2011MNRAS.413.2525G} explored a model for RM observations of kiloparsec scale AGN jet lobes that incorporates both internal and external Faraday depolarisation effects, finding that an amplified, swept-up IGM magnetic field can create band-like fluctuations in the RM. In observations, these bands may be hidden by foreground RM components that follow a Kolmogorov power spectrum. 

This paper continues previous investigations into the RM properties of simulated AGN jets by \citet{2011MNRAS.417..382H,2011MNRAS.418.1621H}. Those authors performed 3D magnetohydrodynamic (MHD) jet simulations with turbulent cluster magnetic fields and found that the compression of the intracluster medium (ICM), particularly by very light jets, enhances the RM. The enhancement is strongest towards the edge of the lobes and hence is expected to impact the statistics of the RM distribution. These studies represent some of the first works studying the RM in simulated AGN jets with magnetic fields evolved self-consistently with the jet, which we add to with this work.

We now aim to break the degeneracy between AGN jet and environment parameters using RMs. Our method uses the RM information for AGN jets as a proxy for the line-of-sight environment. Because the RM depends on the electron density and magnetic field along the line of sight, as shown in Equation \ref{eqn:RM}, the properties of the RMs are expected to differ between AGN jets in environments with different densities and magnetic fields. We test this method by simulating three combinations of jet and environment parameters and then comparing their radio observables and RM maps. We choose jet and environment parameters that closely resemble Cygnus A, as it is the archetypal powerful radio galaxy \citep{1996A&ARv...7....1C}.

The paper is structured as follows. In Section \ref{section:simulations}, we introduce the method and our simulation suite. We describe the implementation of the magnetic field in these simulations in Section \ref{section:B_field_setup}. In Section \ref{section:results} we present the dynamics and synthetic Stokes I emission. We present the RM maps and distributions in Section \ref{section:RM_results} and show that our simulated radio sources can be distinguished using the RM. In Section \ref{section:discussion}, we discuss our findings in the context of current and upcoming observing capabilities and then conclude with a summary of our findings in Section \ref{section:conclusions}.

\section{Simulations}
\label{section:simulations}

The simulations in this study are run with the \textsc{PLUTO} astrophysical fluid dynamics code \citep[version 4.3;][]{2007ApJS..170..228M,2012ApJS..198....7M}, using the relativistic magnetohydrodynamics (RMHD) physics module. The simulation setup used is similar to previous work \citep{2021MNRAS.508.5239Y,2023PASA...40...14Y}, extended to include magnetic fields for both the jet and environment, as described in Section \ref{section:B_field_setup}. The latter follows the approaches described in \citet{2011MNRAS.417..382H} and \citet{2014MNRAS.443.1482H}. We used the HLLD Riemann solver, 2nd-order dimensionally unsplit Runge-Kutta time stepping, and linear reconstruction. The $\nabla \cdot \bm{B} = 0$ condition is controlled by Powell's eight-wave formulation \citep{powell_approximate_1997,1999JCoPh.154..284P}, which is used to minimise numerical artefacts on the simulation grid.

These simulations use PLUTO's passive Lagrangian tracer particles \citep{2018ApJ...865..144V}. To sample the majority of the jet volume, these particles are injected with the jet fluid every 0.01 Myr. They record position, fluid variables, and time since the particle was last shocked \citep[we refer the reader to][for the details on shock flagging]{yates-jones_praise_2022}. These quantities are used to calculate synchrotron emissivities for each calculation using PRAiSE, a modified version of the \textit{Radio AGN in Semi-analytic Environments} \citep[RAiSE; ][]{2018MNRAS.473.4179T} model. This model takes into account spatially resolved adiabatic, synchrotron, and inverse-Compton losses of synchrotron emitting electrons. We refer the reader to \citet{yates-jones_praise_2022} for the full details of the synchrotron emissivity calculation.

These simulations were carried out on a three-dimensional Cartesian grid centred at $(0,0,0)$. Each dimension contains 5 grid patches: a uniform grid from $-2 \rightarrow +2$ kpc with a resolution of $0.04$ kpc/cell; two stretched grid patches from $\pm 2 \rightarrow \pm 10$ kpc; and two stretched grid patches from $\pm 10 \rightarrow \pm 150$ kpc. The high central resolution ensures that the jet injection is sufficiently resolved. The stretched grid patches contain 100 and 150 cells respectively, with typical resolutions of $0.14$ kpc/cell at $10$ kpc and $2.01$ kpc/cell at $100$ kpc. All the simulation grid boundaries are periodic to match our magnetic field initial condition, since the generated environment magnetic fields are periodic by nature. 

The simulations are listed in Table \ref{tab:sim_list}. We perform three simulations, exploring powerful Fanaroff-Riley type II \citep[FR-II;][]{1974MNRAS.167P..31F} sources. The names of each simulation (QX-DY-BZ) correspond to the jet power used as X$\times 10^{38}$ W, central environment density as Y$\times 10^{-26}$ g/cm$^3$, and average environment magnetic field strength as Z $\mu$G. Simulation Q6.5-D4-B1 closely resembles Cygnus A. Simulations Q10.8-D1-B1 and Q6.5-D4-B2 change the jet power, environment density, and average environment magnetic field strength to compare the observational signatures in each case. We have chosen these parameters to explore the jet and environment parameter degeneracy for sources similar to the archetypical Cygnus A.

We use an average cluster magnetic field strength of $1 \mu$G in simulations Q6.5-D4-B1 and Q10.8-D1-B1, and $2 \mu$G for simulation Q6.5-D4-B2. The Cygnus A cluster magnetic field is estimated to be $\sim 5 \mu$G \citep{1996A&ARv...7....1C} on average. The lower magnetic field strengths are more representative of a typical FR-II radio galaxy environment \citep{2002ARA&A..40..319C}. This lower magnetic field also assists with numerical stability in the less dense environment of simulation Q10.8-D1-B1. In this lower-density environment, a $5 \mu$G magnetic field would be dynamically important. Combined with the diffusion of the magnetic field, such a strong field would alter the density and pressure of the environment and introduce transonic flows on the simulation grid.

Our simulations were run using the \textit{kunanyi} high performance computing facility provided by Digital Research Services, IT Services at the University of Tasmania. Each simulation ran on 1680 Intel cores and took an average CPU time of 320,000 hours.

\subsection{Environment}

\begin{figure*}
    \centering
    \includegraphics[width=\textwidth]{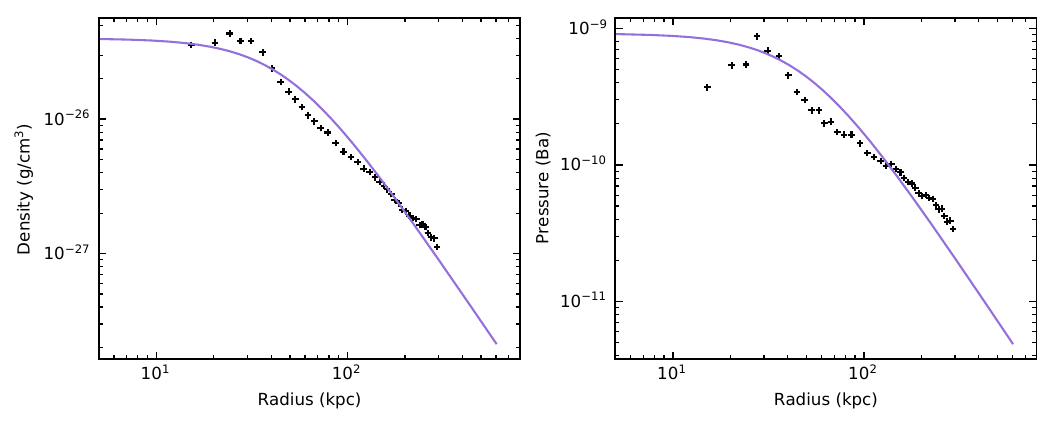}
    \caption{Density (left) and pressure (right) profiles of the Cygnus A cluster. Black crosses indicate data points from Fig. 4 of \citet{2018ApJ...855...71S}. The data is fitted with an isothermal King profile with temperature $1.4 \times 10^8$ K to match the density and pressure on scales of 50-150 kpc.}
    \label{fig:CygA_tabenv}
\end{figure*}

The jets are simulated in an idealised environment based on \textit{Chandra} data of the Cygnus A cluster. \citet{2018ApJ...855...71S} use 2.0 Msec of X-ray observations to derive pressure, density, and temperature profiles as shown in their Fig. 4. The density and pressure profiles are fitted to an isothermal King profile of the form: 

\begin{equation}
    \rho = \rho_0 \left(1+ \left(\tfrac{r}{r_c}\right)^2 \right)^{-3 \beta / 2},
\end{equation}

shown in Fig. \ref{fig:CygA_tabenv}. The fitted Cygnus A profile has $r_c = 50$ kpc, $\beta = 0.885$, $\rho_0 = 4 \times 10^{-23}$ kg/m$^3$, and $T_0 = 1.4 \times 10^8$ K. To create a lower-density environment for simulation Q10.8-D1-B1, the shape of the cluster profile is kept but $\rho_0$ is reduced by a factor of $4$. Temperature is held constant in all simulations. Simulation Q6.5-D4-B2 uses the same cluster density and pressure as simulation Q6.5-D4-B1.

\subsection{Jet parameters}
\label{section:jet_params}

The primary jet parameters in our simulations are the kinetic power, speed, and half-opening angle. The jet is injected with a Lorentz factor of 5, corresponding to a speed of $0.995 c$. This is consistent with evidence that FR-II sources, including Cygnus A, have initially relativistic jets which drive energy out to kpc scales \citep{1996A&ARv...7....1C,hardcastle_radio_2020}. VLBI measurements estimate the full jet opening angle for Cygnus A to be $\sim 10\degree$ on parsec scales \citep{2016A&A...585A..33B,2014evn..confE..16B}. To properly resolve jet injection on kiloparsec scales, we use a half-opening angle of $15 \degree$. The jet power in simulations Q6.5-D4-B1 and Q6.5-D4-B2 is $6.5 \times 10^{38}$ W, as estimated from X-ray observations of the bow shock surrounding Cygnus A (\citeauthor{2018ApJ...855...71S} \citeyear{2018ApJ...855...71S}; see also \citeauthor{1999MNRAS.305..707K} \citeyear{1999MNRAS.305..707K}). A higher jet power of $10.8 \times 10^{38}$ W is used in simulation Q10.8-D1-B1. In Section \ref{section:radio}, we show that the combination of the higher jet power and lower central density in simulation Q10.8-D1-B1 produces size-luminosity tracks similar to simulation Q6.5-D4-B1. We use a fluid tracer to quantify jet-environment mixing. Jet material is injected with a tracer value of 1, while the environment has an initial value of 0.

\begin{table}
	\centering
	\caption{Parameters of the three simulations.  $Q_{\rm jet}$ is the one-sided kinetic jet power of the injected jet. $\rho_{0, {\rm env}}$ is the central density of the environment, and $\Bar{B}_{\rm env}$ is the average environment magnetic field strength.}
	\label{tab:sim_list}
    \begin{tabular}{cccc}
    \hline
         Name & $Q_{\rm jet}$ (W) & $\rho_{0, {\rm env}}$ (g/cm$^3$) & $\Bar{B}_{\rm env}$ ($\mu$G)  \\
    \hline
         Q6.5-D4-B1 & $6.5 \times 10^{38}$ & $4 \times 10^{-26}$ & 1 \\
         Q10.8-D1-B1 & $10.8 \times 10^{38}$ & $1 \times 10^{-26}$ & 1  \\
         Q6.5-D4-B2 & $6.5 \times 10^{38}$ & $4 \times 10^{-26}$ & 2  \\
    \hline
    \end{tabular}
\end{table}

\section{Magnetic field setup}
\label{section:B_field_setup}

\subsection{Jet magnetic field}
\label{section:B_field_jet}

We extend the hydrodynamical setup of \citet{2021MNRAS.508.5239Y} as follows. The helical magnetic field in the jet is approximated as toroidal field loops in the jet injection region on the simulation grid. This approximation is valid since the toroidal field dominates the poloidal field in the jet collimation region \citep{2012SSRv..169...27P}. Our toroidal field loops are parameterised in the following way:

\begin{equation}
    \begin{bmatrix}
    B_x \\
    B_y \\
    B_z
    \end{bmatrix} =
    \begin{bmatrix}
    -\frac{B_{0}\sin(\phi) \left(x^2 + y^2\right)^{\frac{1}{2}}}{r} \\
    \frac{B_{0}\cos(\phi) \left(x^2 + y^2\right)^{\frac{1}{2}}}{r} \\
    0
    \end{bmatrix},
\end{equation}

where $B_0$ is the amplitude of the magnetic field in the jet, $\phi$ is the spherical azimuth angle, $x$ and $y$ are the distances in their respective axes from the jet ($z$) axis, and $r$ is the distance from the origin along the outside of the jet cone. This setup ensures numerical stability in the injection region. Only the magnetic field at the outer boundary of the injection region defines the simulated jets. These components of the magnetic field are directly assigned to the PLUTO magnetic field variables when the jet is injected into the simulation. The initial value of the jet magnetic field $B_0$ is chosen to be $0.15 \mu$G, following \citet{2014MNRAS.443.1482H}.

\subsection{Environment magnetic field}
\label{section:B_field_env}

We follow the methods of \citet{murgia_magnetic_2004} and \citet{2013MNRAS.433.3364H} to generate turbulent magnetic fields in the cluster gas, using a Kolmogorov power spectrum with slope $\zeta = 17/3$. The magnetic field is set up by first generating the magnetic vector potential components in Fourier space. The magnitudes of these vector components are randomly selected from a Rayleigh distribution with variance $|A_k|^2$, which is given by:

\begin{equation}
    |A_k|^2 = \left(k_x^2 + k_y^2 + k_z^2\right)^{-\zeta}.
\end{equation}

Here, $k_x$, $k_y$, and $k_z$ are the wavenumbers in the x, y, and z coordinates. These correspond to the discrete Fourier transform sample frequencies generated with a window length of $600$, which matches the number of cells in each dimension on the simulation grid. Next, three amplitude arrays $A_i$ are generated, one for each component of the magnetic vector potential ($i = x, y, z$), which have the same dimensions as the simulation grid. Then, the phase arrays of each magnetic vector component, $\phi_i$, have phases randomly drawn from a uniform distribution between $0$ and $2 \pi$. These phase arrays are then combined with the amplitude arrays to generate the magnetic vector potential, as $\bm{A}_i = A_i e^{j \phi_i}$ for $i = x, y, z$. The Fourier transform of the magnetic field is then given by:

\begin{equation}
\label{eqn:curl}
    \bm{B}_i = j \bm{k} \times \bm{A}_i,
\end{equation}

where $\bm{k}$ is the wavenumber vector in three dimensions, $\bm{A}_i$ are the magnetic vector potential components in Fourier space, and $\bm{B}_i$ are the magnetic field components in Fourier space. An inverse Fourier transform is performed to find the real values of the magnetic field strength vector components $B_x$, $B_y$, and $B_z$ at each point on the given simulation grid. These dimensionless magnetic field vectors are then scaled to realistic galaxy cluster values. A scaling array $C$ and scaling constant $D$ are calculated in CGS units as follows:

\begin{equation}
   C =  \left(8 \pi p \right)^{1/2} \: \: ; \: \:
    D = \frac{\Bar{B}_{\rm env}}{\langle C(B_x^2 + B_y^2 + B_z^2)^{1/2}\rangle},
\end{equation}

where $\Bar{B}_{\rm env}$ is the average environment magnetic field strength as shown in Table \ref{tab:sim_list} and $p$ is the environment pressure. The dimensionless magnetic field vectors are multiplied by both $C$ and $D$ to generate the final cluster magnetic field. This is slightly different to the method described in \citet{2011MNRAS.418.1621H}, where the authors scale the magnetic vector potential $\bm{A}_i$ to the radial density profile before taking the curl as in Equation \ref{eqn:curl}. Our method may introduce some small errors in $\nabla \cdot \bm{B}$, which is removed by PLUTO's divergence correction algorithm.

\begin{figure}
    \centering
    \includegraphics[width=\columnwidth]{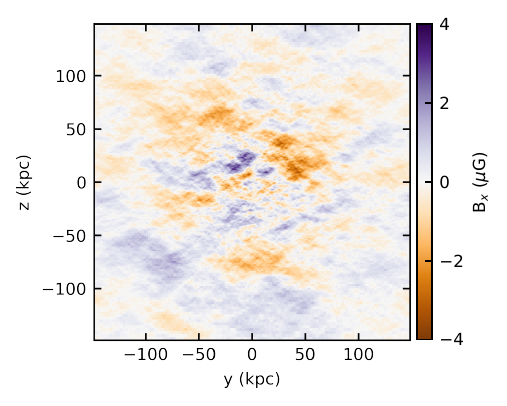}    
    \caption{Midplane slice at $x=0$ of the $x$-component of the initial turbulent magnetic field generated for simulation Q6.5-D4-B1.}
    \label{fig:generated_B_field}
\end{figure}

The ratio of the average magnetic field energy density to the average pressure of the environment is calculated to ensure that the magnetic field is less than $10\%$ of the thermal pressure, consistent with expectations for cluster magnetic fields that are not dynamically dominant \citep{2002ARA&A..40..319C}. This results in an average environment magnetic field strength on the order of a few $\mu$G, which is typical for clusters \citep{2002ARA&A..40..319C}. This condition is applied when generating the environment magnetic field to be loaded into the simulation as an initial condition. The generated initial magnetic field for these simulations is shown in Fig. \ref{fig:generated_B_field}.

\section{Results}
\label{section:results}

\subsection{Dynamics and implications for the RM results}
\label{section:dynamics}

\begin{figure*}
    \centering
    \includegraphics[width=\textwidth]{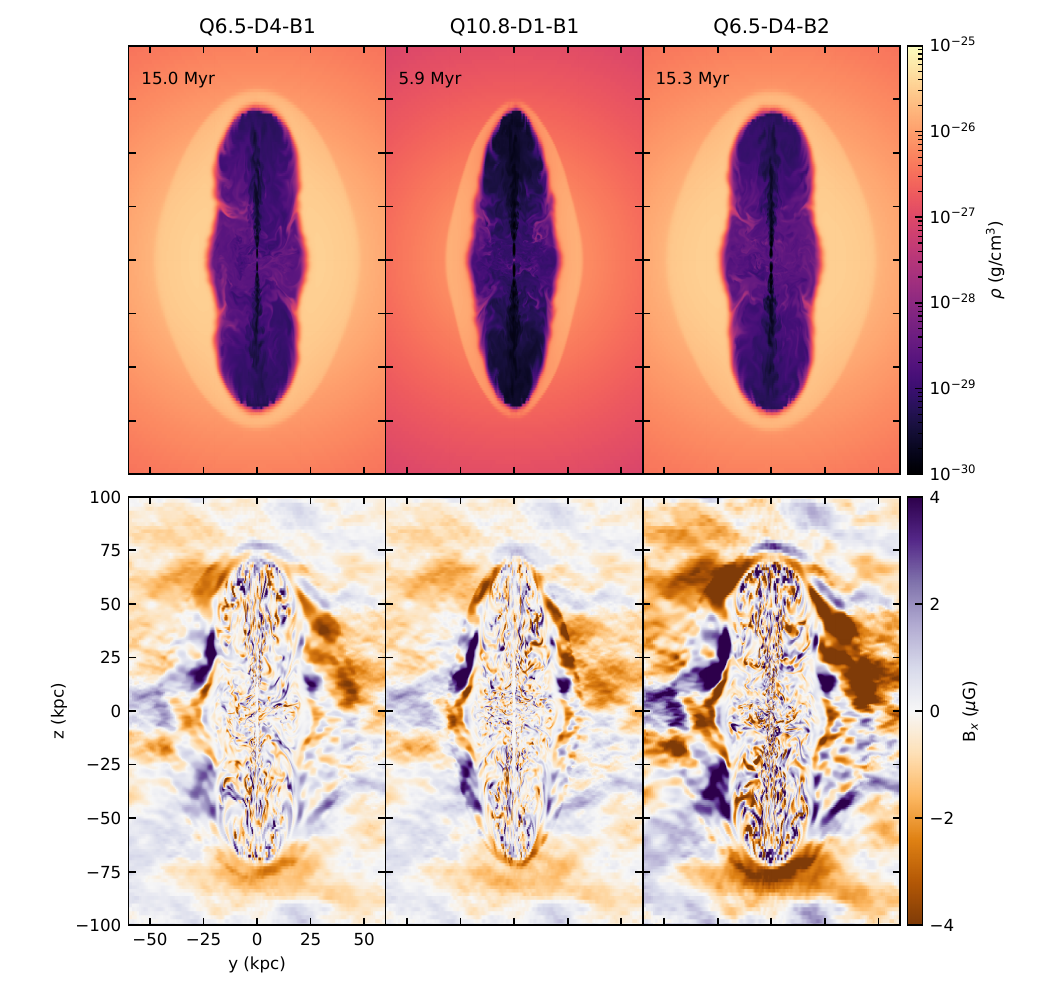}
    \caption{Midplane slices at $x=0$ of density (top) and $x-$component of the magnetic field ($B_x$; bottom) for simulations Q6.5-D4-B1 (left), Q10.8-D1-B1 (centre), and Q6.5-D4-B2 (right). Simulation Q6.5-D4-B1 is plotted at 15.0 Myr, simulation Q10.8-D1-B1 is plotted at 5.9 Myr, and simulation Q6.5-D4-B2 is plotted at 15.3 Myr.}
    \label{fig:dynamics_rho_B}
\end{figure*}

To show the morphological differences between the simulations, we plot $y-z$ midplane slices of the density and $x$-component of the magnetic field in Fig. \ref{fig:dynamics_rho_B}. Simulation Q6.5-D4-B1 is shown at $t = 15.0$ Myr, simulation Q10.8-D1-B1 is shown at $t = 5.9$ Myr, and simulation Q6.5-D4-B2 is shown at $t = 15.3$ Myr. These times are chosen such that the total extent of their radio emission matches the observed length of Cygnus A in the sky \citep[$141.5$ kpc;][]{2002ARA&A..40..319C,2019MNRAS.486.1225T}. The details of the length calculation are discussed in Section \ref{section:radio}. 

The morphology of the sources in simulations Q6.5-D4-B1 and Q6.5-D4-B2 are almost identical, as expected, because the jet power and environment density are the same and the cluster magnetic field is not dynamically significant (Section \ref{section:B_field_env}). The morphological differences between these simulations and simulation Q10.8-D1-B1 correspond to the difference in jet power: the jet power in simulation Q10.8-D1-B1 is 1.66 times higher than that in simulations Q6.5-D4-B1 and Q6.5-D4-B2. Due to the higher jet power and lower environment density in simulation Q10.8-D1-B1, the jet propagates faster. This results in narrower lobes as the backflow has had less time to fill them out \citep[see e.g.][]{2023PASA...40...14Y}. Simulations Q6.5-D4-B1 and Q6.5-D4-B2 have a much slower expansion in the jet direction and wider lobes, and so the sources in these simulations are about 2.5 times older than the source in simulation Q10.8-D1-B1 for the same jet length. Despite the difference in jet expansion speeds, all three simulations have the same central ($z=0$) lobe width at the snapshots pictured in Fig. \ref{fig:dynamics_rho_B}.

We find that the average Mach number of the bow shock (using the furthest extent of the bow shock along the jet axis from the origin divided by the source age) is $2.8, 6.6,$ and $2.8$ for simulations Q6.5-D4-B1, Q10.8-D1-B1, and Q6.5-D4-B2, respectively. In contrast, the instantaneous Mach number is $2.48, 4.90,$ and $2.02$, respectively. The lower Mach numbers for simulations Q6.5-D4-B1 and Q6.5-D4-B2 indicate that the bow shocks of the jets in those simulations are weaker than in simulation Q10.8-D1-B1.

The extent of the bow shock defines the region of the environment affected by the jet. As the jet expands supersonically into the cluster environment, its bow shock sweeps up ambient gas into a `shocked shell' between the bow shock and the low-density jet cocoon \citep{2002MNRAS.335..610A}. Simulations by \citet{2011MNRAS.417..382H,2011MNRAS.418.1621H} have shown that the cluster magnetic field is compressed and stretched during jet cocoon expansion, amplifying the strength of the magnetic field. The bow shock initially expands self-similarly \citep{2002MNRAS.335..610A}, but the transverse expansion slows as the jet cocoon approaches pressure equilibrium with the environment \citep{gaibler_very_2009,1997MNRAS.292..723K}. These shocked shells are dependent on the interaction between the jet and its environment.

The differences between simulations Q6.5-D4-B1/Q6.5-D4-B2 and Q10.8-D1-B1 are clearly shown by the shape and thickness of the shocked shell. The shocked shells in the lower-powered jet simulations are both thicker and rounder. These jets are older than the higher-powered jet for the same length, so the bow shocks in simulations Q6.5-D4-B1 and Q6.5-D4-B2 have had more time to expand laterally. In these simulations, the thickness of the shell at $z=0$ increases linearly over time at a rate of $\sim 1.8$ kpc/Myr, whereas in simulation Q10.8-D1-B1 the shell thickness increases at a rate of $\sim 1.5$ kpc/Myr. The jet cocoons are still overpressured compared to the ambient environment, resulting in this linear relationship. As these jets continue to evolve, the expansion of the shocked shell of the lower-powered jets will begin to slow down. This difference in the rate of shell growth indicates that at an age of $\sim 15$ Myr, the shell in simulation Q10.8-D1-B1 would still be thinner than in simulations Q6.5-D4-B1 and Q6.5-D4-B2. Additionally, the source would be much larger, so the size-luminosity signature would be quite different in this case. Since the shocked shells in simulations Q6.5-D4-B1 and Q6.5-D4-B2 are significantly thicker than in simulation Q10.8-D1-B1, this will influence the strength of the Faraday rotation signal, as shown below.

The shocked shell is also a region of enhanced density and magnetic field strength. Therefore, it will have high RM values (Equation \ref{eqn:RM}). Following \citet{2011MNRAS.418.1621H}, we identify the shocked shell as the Faraday screen. A greater Faraday screen thickness corresponds to a greater path length through this amplified density and magnetic field region (Equation \ref{eqn:RM}). The median path length through the Faraday screen looking down the negative $x-$axis is $27.8$ kpc for simulation Q6.5-D4-B1, $11.9$ kpc for simulation Q10.8-D1-B1, and $28.5$ kpc for simulation Q6.5-D4-B2. Similar median path lengths in simulations Q6.5-D4-B1 and Q6.5-D4-B2 are expected, as the dynamics of these two simulations are almost identical. Since simulations Q6.5-D4-B1 and Q6.5-D4-B2 have a higher environment density and Faraday screen thickness than simulation Q10.8-D1-B1, we expect that the rotation measure values from the Faraday screen in these simulations will be higher.

We find that the difference in the density values in the Faraday screen between simulations Q6.5-D4-B1/Q6.5-D4-B2 and Q10.8-D1-B1 is less pronounced than the difference in environment densities. The higher average and instantaneous Mach number of the bow shock in simulation Q10.8-D1-B1 leads to greater compression and a more effective increase in the gas density. This is shown by the higher mean density amplification ratio (Table \ref{tab:amplification_ratios}). However, due to the lower ambient density, the mean density in the Faraday screen of simulation Q10.8-D1-B1 is still less than half that of the Faraday screens in simulations Q6.5-D4-B1 and Q6.5-D4-B2 (Table \ref{tab:amplification_ratios}). Therefore, the Faraday rotation signal from simulation Q10.8-D1-B1 is expected to be lower than that of simulations Q6.5-D4-B1 and Q6.5-D4-B2.

\begin{table}
    \centering
    \caption{Top to bottom: Faraday screen values of; mean density amplification ratio, mean density, median magnitude of the $x-$component of the magnetic field, volume, mass and mass percentage. The mass percentage is taken as the Faraday screen mass divided by the total mass within 150 kpc at $t=0$ Myr. These values are taken at $15.0, 5.9,$ and $15.3$ Myr for simulations Q6.5-D4-B1, Q10.8-D1-B1, and Q6.5-D4-B2, respectively. $\rho_e$ is the density within the region of the Faraday screen taken at the nearest time in an environment-only run of each simulation (e.g. 15 Myr, 6 Myr, and 15 Myr, respectively).}
    \begin{tabular}{c|ccc}
    \hline
        Simulation Name & Q6.5-D4-B1 & Q10.8-D1-B1 & Q6.5-D4-B2  \\
        \hline
         Mean $\dfrac{\rho}{\rho_e}$ & 1.4 $\pm$ 0.2 & 1.9 $\pm$ 0.4 & 1.3 $\pm$ 0.2 \\
         Mean $\rho$ ($10^{-26}$ g/cm$^3$) & 2.9 $\pm$ 0.4 & 1.2 $\pm$ 0.2 & 2.7 $\pm$ 0.4 \\
         Median $|B_{x}|$ ($\mu$G) & 0.8 $\pm$ 0.9 & 1.1 $\pm$ 1.2 & 1.7 $\pm$ 1.7 \\
         Volume (kpc$^3$) & $4.96 \times 10^5$ & $1.32 \times 10^5$ & $5.43 \times 10^5$ \\
         Mass ($M_\odot$) & $1.96 \times 10^{11}$ & $2.28 \times 10^{10}$ & $2.08 \times 10^{11}$ \\
         $\dfrac{M_{\rm screen}}{M_{150}}$ & 19.3\% & 8.96\% & 20.4\% \\
         \hline
    \end{tabular}
    \label{tab:amplification_ratios}
\end{table}

The bow shock driven by the jet also amplifies and sweeps out the magnetic field. Simulation Q6.5-D4-B2 has an ambient magnetic field that is twice as strong as in simulations Q6.5-D4-B1 and Q10.8-D1-B1. This difference increases the median $x-$component magnetic field value in the Faraday screen, which agrees with the findings of \citet{2011MNRAS.417..382H,2011MNRAS.418.1621H}. Therefore, we expect the Faraday rotation measure values in simulation Q6.5-D4-B2 to be higher than in simulations Q6.5-D4-B1 and Q10.8-D1-B1.

\subsection{Radio emission at 151 MHz}
\label{section:radio}

\begin{figure*}
    \centering
    \includegraphics[width=\textwidth]{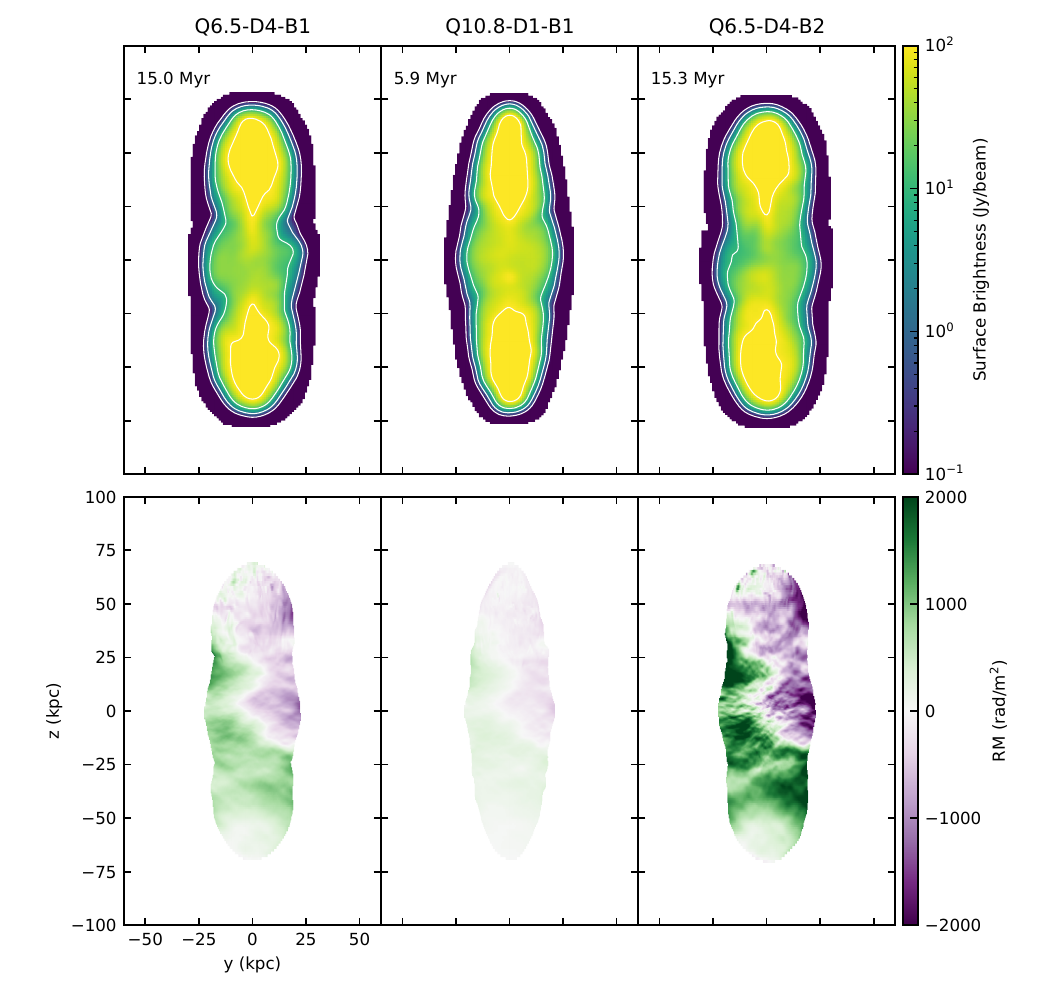}
    \caption{Top: Surface brightness at 151 MHz for simulations Q6.5-D4-B1, Q10.8-D1-B1, and Q6.5-D4-B2 (left, centre, and right). Contours are shown for 0.1, 1, 10, 100 Jy/beam. We take a beam size of $4.45$ arcsec. All sources have been tilted by $15 \degree$ out of the page, where the upper jet is tilted towards the reader. Bottom: Rotation measure integrated along the (tilted) positive x-axis. Simulation Q6.5-D4-B1 is plotted at 15.0 Myr, simulation Q10.8-D1-B1 is plotted at 5.9 Myr, and simulation Q6.5-D4-B2 is plotted at 15.3 Myr.}
    \label{fig:radio_sb_RM}
\end{figure*}

In the top row of Fig. \ref{fig:radio_sb_RM} we show synthetic surface brightness images of the three simulations at 151 MHz. This synthetic emission is generated using the PRAiSE code \citep{yates-jones_praise_2022}, which is based on the analytical modelling work by \citet{2018MNRAS.473.4179T} (for a detailed description of the mathematics and computational implementation of this modelling, the reader is directed to these works). We adopt parameters consistent with Cygnus A, placing our simulated sources at a redshift of $0.056075$ \citep{1997ApJ...488L..15O} and at a  $15 \degree$ (toward the reader) viewing angle \citep{2016A&A...585A..33B}. We used minimum and maximum Lorentz factors $\gamma_{\rm min} = 600$ and $\gamma_{\rm max} = 1.3 \times 10^5$ \citep{2016MNRAS.463.3143M}, an equipartition factor $\eta = 0.12$, and an injection index $\alpha_{\rm inj} = 0.7$ \citep[to approximate the 151 MHz - 327.5 MHz spectral index;][]{2010MNRAS.401...67S}. Our beam FWHM is $4.45$ arcsec. Here, we define the spectral index as $S_\nu \propto \nu^{-\alpha}$.

We plot each simulation at the time where the total projected lobe length corresponds to the observed extent of Cygnus A, $141.5$ kpc. Following \citet{2023PASA...40...14Y}, we calculate these projected lengths from the surface brightness as the distance from the jet injection point to the furthest point of emission, two orders of magnitude below the maximum surface brightness. The total length is the sum of the jet and counter-jet lengths.

At these snapshots, the 151 MHz morphology of each simulation is quite similar. Each simulation exhibits FR-II-like morphology of edge-brightened lobes, as well as hotspot-like features. The lobes of simulations Q6.5-D4-B1 and Q6.5-D4-B2 are more “pinched” than simulation Q10.8-D1-B1's lobes, with less emission in the central regions. The underlying properties of these sources would be difficult to distinguish based upon their appearance alone.

\begin{figure}
    \centering
    \includegraphics[width=\columnwidth]{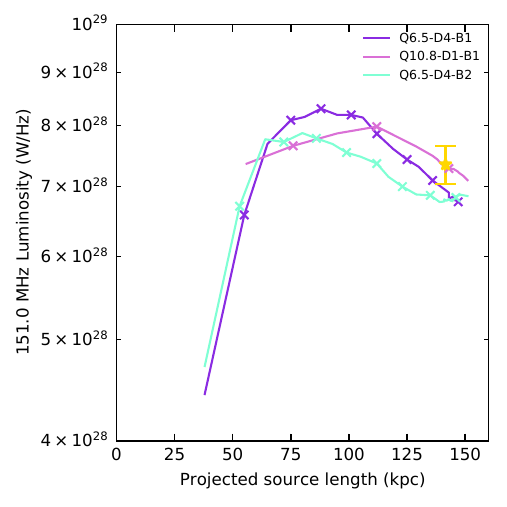}
    \caption{Size-luminosity tracks for simulations Q6.5-D4-B1, Q10.8-D1-B1, and Q6.5-D4-B2 at 151 MHz. The gold star corresponds to Cygnus A, with a total projected length of 141.5 kpc \citep{2019MNRAS.486.1225T} and 151 MHz radio luminosity of $7.3 \pm 0.3 \times 10^{28}$ W/Hz \citep{1977A&A....61...99B}. Crosses indicate projected lobe lengths every 2 Myr. Simulations Q6.5-D4-B1, Q10.8-D1-B1, and Q6.5-D4-B2 meet the projected Cygnus A length at 15.0, 5.9, and 15.3 Myr, respectively.}
    \label{fig:PD_tracks}
\end{figure}

We plot the size-luminosity tracks of each source in Fig. \ref{fig:PD_tracks}. The tracks for each simulation are similar to each other and broadly consistent with the 151 MHz luminosity of Cygnus A at a projected length of 141.5 kpc. The derived ages of each simulation, $t = 15.0$ Myr, $t = 5.9$ Myr, and $t = 15.3$ Myr, are also broadly consistent with Cygnus A age estimates of 6 - 30 Myr \citep{1991ApJ...383..554C,1996A&ARv...7....1C, 1999MNRAS.305..707K}. This consistency indicates that the simulation parameters chosen are acceptable for comparison to powerful radio galaxies. The luminosity at a given size for the three simulations remain within a range of $\pm 15 \%$ of each other after a projected length of $70$ kpc is reached. Without extra information about the environment, these sources are effectively identical to one another based on their size-luminosity tracks alone. In Section \ref{section:RM_results}, we show that these sources can be distinguished using Faraday rotation measure information.

\subsection{Faraday rotation}
\label{section:RM_results}

The Faraday rotation measures are calculated directly from the simulation fluid variables. We integrate Equation \ref{eqn:RM} through lines of sight that pass through jet material, calculating two separate RM components: the undisturbed external medium, and the Faraday screen. Here, we take lines of sight down the $x$-axis to calculate spatially resolved maps in the $y-z$ plane of the simulation grid. The external RM component is calculated from the edge of the simulation grid (i.e. from the face of a cube) to the edge of the Faraday screen. We distinguish between the jet and ambient media using a fluid tracer threshold of $10^{-4}$ \citep[e.g.][]{2013MNRAS.430..174H,2018MNRAS.480.5286Y}.

The bottom row of Fig. \ref{fig:radio_sb_RM} displays the RM maps for each simulation at the same snapshots as in Fig. \ref{fig:dynamics_rho_B}. The jet axis is tilted out of the page by 15 degrees for each simulation. Although the initial magnetic field structure is identical across all simulations (and hence the broad structure of the RM is also the same), there are significant differences in the \textit{magnitude} of the RM maps. The range of total RM values reached in simulations Q6.5-D4-B1, Q10.8-D1-B1, and Q6.5-D4-B2 is (-1518, 1678), (-508, 744), and (-3005, 3171) rad/m$^2$, respectively. Recent polarisation observations of Cygnus A from \citet{2020ApJ...903...36S} found rotation measures ranging between -4500 to +6400 rad/m$^2$ in the eastern lobe and -5000 to +3000 rad/m$^2$ in the western lobe (corresponding to our bottom and top lobes respectively); our simulation Q6.5-D4-B2 is most similar to these Cygnus A values.

\begin{figure}
    \centering
    \includegraphics[width=\columnwidth]{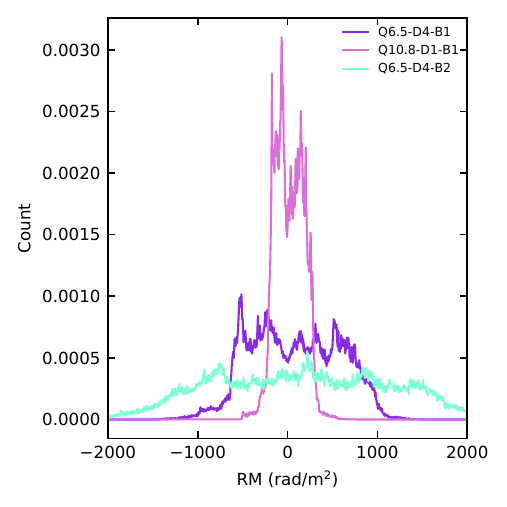}
    \caption{RM distributions for simulations Q6.5-D4-B1, Q10.8-D1-B1, and Q6.5-D4-B2. All jets have been tilted to a viewing angle of $15 \degree$, where the upper jet is tilted out of the page. The RM is integrated down the (tilted) positive $x$-axis. The jet is oriented along the $z$-axis.}
    \label{fig:RM_hist_allsims}
\end{figure}

To demonstrate the differences in RM between the three simulations, we plot the distribution of rotation measure values for each simulation in Fig. \ref{fig:RM_hist_allsims}. Each distribution has a centre close to zero, but they have different widths. The mean and standard deviation for simulations Q6.5-D4-B1, Q10.8-D1-B1, and Q6.5-D4-B2 are $79 \pm 488, 19 \pm 156,$ and $140 \pm 970$ rad/m$^2$, respectively. We use the full-width at half-maximum (FWHM) of the RM distribution to measure the differences between our simulations. We calculate FWHMs of $1430 \pm 9, 474 \pm 11,$ and $2983 \pm 56$ rad/m$^2$ in simulations Q6.5-D4-B1, Q10.8-D1-B1, and Q6.5-D4-B2 respectively.

Comparing simulation Q6.5-D4-B1 to simulation Q10.8-D1-B1, we see that increasing the environment density by a factor of 4 broadens the distribution by a factor of $\sim3$. Furthermore, the increase in the distribution width from simulation Q6.5-D4-B1 to simulation Q6.5-D4-B2 shows us that increasing the magnetic field strength by a factor of $2$ also broadens the distribution by a factor of $\sim2$. In both cases, these environmental increases amplify the extreme values of the RM. Therefore, we have shown that we can distinguish between our simulations with similar Stokes I properties (Figs \ref{fig:radio_sb_RM} and \ref{fig:PD_tracks}) using the RM values.

\begin{figure*}
    \centering
    \includegraphics[width=\textwidth]{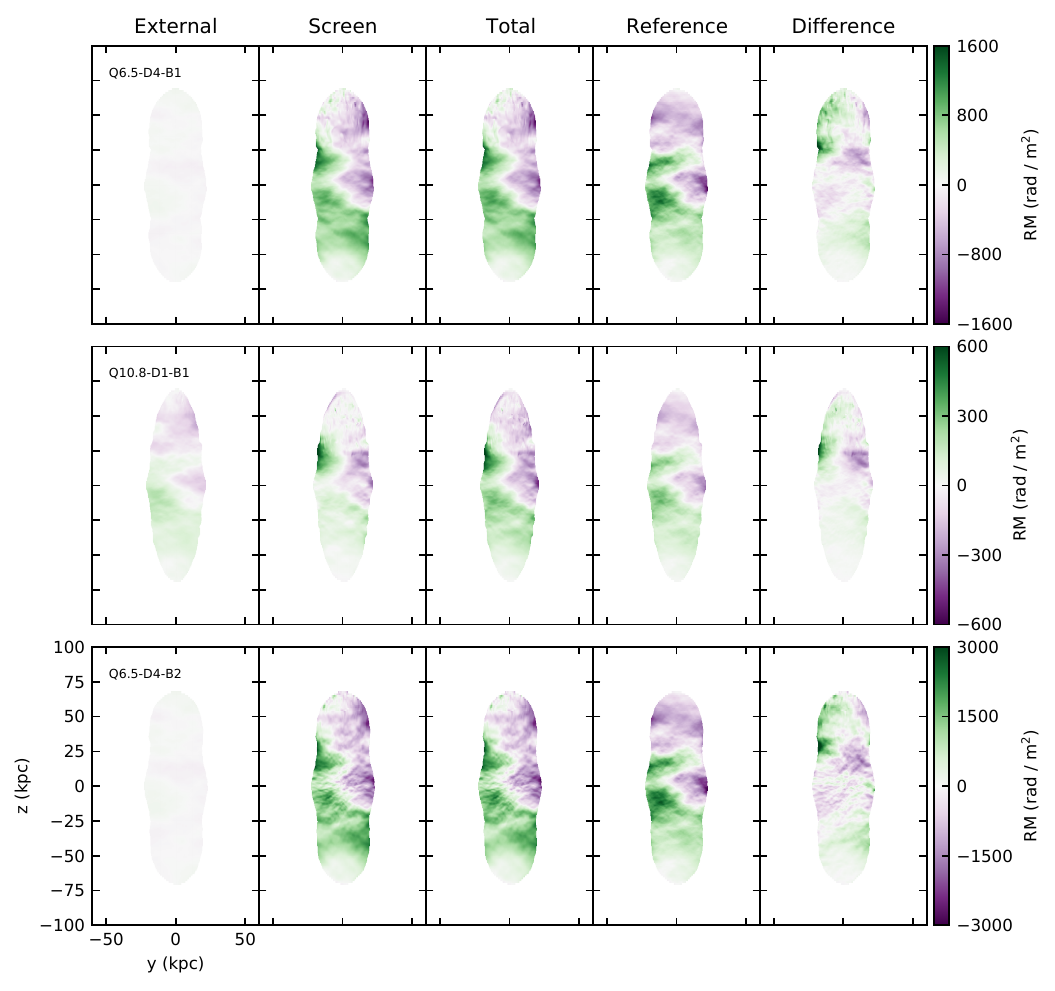}
    \caption{Spatial distributions of the separated components of the rotation measure. From left to right, we plot the external component, Faraday screen component, and the total sum of these two. In the fourth column, we plot the environment RM at $t = 0$ Myr using the jet tracer from each simulation's time, respectively. The final column is the difference between the total RM and the environment RM. As in Fig. \ref{fig:radio_sb_RM}, RM is integrated along the (tilted) positive $x$-axis. The jet is oriented along the $z$-axis and is tilted at a viewing angle of $15\degree$ out of the page, where the upper jet is tilted towards the reader. From top to bottom, we plot simulation Q6.5-D4-B1 at 15.0 Myr, simulation Q10.8-D1-B1 at 5.9 Myr, and simulation Q6.5-D4-B2 at 15.3 Myr.}
    \label{fig:RM_components}
\end{figure*}

\subsubsection{RM as a probe of environment}
\label{section:RM_probe_env}

The distinction between the Faraday rotation measures allows us to break the degeneracy in the observable radio properties of our three simulations, using the Faraday rotation as a probe of the radio galaxy environment.

We plot the components of the RM for our three simulations in Fig. \ref{fig:RM_components}. The first two columns correspond to the two components (external and Faraday screen) that make up the total RM (third column) for our simulations. The fourth column corresponds to the `reference' RM at $t = 0$ Myr using the same integration limits as for the total RM; in practice, this is calculated by applying a jet cocoon mask to the undisturbed environment, by only considering Faraday rotation at time $t = 0$ Myr by the gas which is external to the jet cocoon at the age of interest. The fifth column (`difference' RM) corresponds to the difference between the total RM and the reference RM. We ignore the internal Faraday rotation component in this case, as we find that the RM in the low-density cocoon is negligible and the amount of entrained thermal plasma is low. We note that the external RM component does not include any chance line-of-sight sources that may be present in real observations.

The dominant component of the total RM in all cases is the Faraday screen. However, in simulation Q10.8-D1-B1, the external RM is more comparable to the Faraday screen RM. In this case, the Faraday screen is a factor of $\sim 2$ thinner than the screens in simulations Q6.5-D4-B1 and Q6.5-D4-B2, and the mean density is lower (Section \ref{section:dynamics}). Additionally, the external RM component in this simulation is higher in magnitude than for simulations Q6.5-D4-B1 and Q6.5-D4-B2, as it probes further into the cluster where the density and magnetic field profiles both increase.

We see that the overall magnitude of the RM is not markedly different when a radio source is present. This is consistent with the findings of \citet{2011MNRAS.418.1621H}, who primarily find RM enhancements towards the edges of the visible lobe. As the jet pushes out the magnetic field, it aligns the field with the shape of the cocoon. At the edges, this magnetic field alignment will be closer to the line of sight of the RM calculation (see Fig \ref{fig:dynamics_rho_B}). These edge enhancements can be seen in the `difference' RMs in Fig. \ref{fig:RM_components}. Edge-enhanced RMs have also been reported in recent observations of the Spiderweb Protocluster \citep{anderson_spiderweb_2022}.

Our rotation measure maps (Figs \ref{fig:radio_sb_RM}, \ref{fig:RM_components}) also show a clear RM sign reversal across the jet (i.e. in the $y-$direction). The RM in the upper lobe has a clear divide between positive and negative RMs on either side of the jet ($z-$)axis. Reversals not aligned with the jet axis can be seen in the `reference' RM maps (i.e. without a jet present), which is expected from the presence of the larger scale modes in the assumed power spectrum. However, we also see in the Faraday screen and total RMs that the jet acts to align this reversal with the jet axis, particularly in the upper lobes ($+z$), which are tilted towards the reader. This is due to the action of the jet compressing and amplifying the ambient magnetic field in the Faraday screen, inducing a change in the field distribution. A similar reversal can also be seen in the western lobe of Cygnus A \citep[Fig. 9 of][]{2020ApJ...903...36S} and the Spiderweb Protocluster \citep{anderson_spiderweb_2022}. 

The lower lobes ($-z$) are tilted away from the reader and have lower values of RM at the tip. This does not match the expectation of the Laing-Garrington effect \citep{1988Natur.331..147G,1988Natur.331..149L}, which predicts that the radio lobe pointed further away from the observer will have greater RM values and therefore be more subject to depolarisation. This lobe will have a greater path length through the Faraday screen, which will act to increase the magnitude of the rotation measure. However, the prominent large-scale field fluctuations in a Kolmogorov power spectrum can lead to RM distributions that differ from the Laing-Garrington expectation. Anisotropic fields can be found when the observing window is similar to the correlation length of the magnetic field \citep{2003A&A...401..835E}. This correlation length is calculated by \citep{2019MNRAS.490.4841L,2003A&A...401..835E}:

\begin{equation}
    \Lambda_\mathit{B}= \frac{3 \pi}{2} \frac{\int^\infty_0 |\mathit{B_k}|^2 \mathit{k} dk}{\int^\infty_0 |\mathit{B_k}|^2 \mathit{k}^2 dk},
\end{equation}

where $\mathit{k}$ is the wavenumber, and $|\mathit{B_k}|^2$ is the Fourier transform of the magnitude of the magnetic field. We find that the correlation length of the field in our simulations is $\sim 220$ kpc. Since the total length of our jets is $141.5$ kpc, the field within the jet observing window will not necessarily be isotropic, which results in the Laing-Garrington effect not being observed in these simulations. This correlation length is larger than the $\sim 30$ kpc coherence length estimated for Cygnus A \citep[][]{2002ARA&A..40..319C}. In our simulations, due to the stretched grid, larger magnetic field clouds were more stable to magnetic diffusion over time. 

Since both the external and Faraday screen RMs are dependent on the environmental magnetic field, modelling this field accurately is important for making comparisons to RM observations. Not only does the seed magnetic field make a difference to the spatial RM distribution, but the line of sight chosen for the RM integration (e.g. $x-$axis or $y-$axis) can also make a significant difference, especially in the case of an anisotropic field as we have here. This aspect of modelling cluster magnetic fields is an area for improvements to be made in the future.

\subsubsection{Observability of the RM differences}

\begin{figure*}
    \centering
    \includegraphics[width=\textwidth]{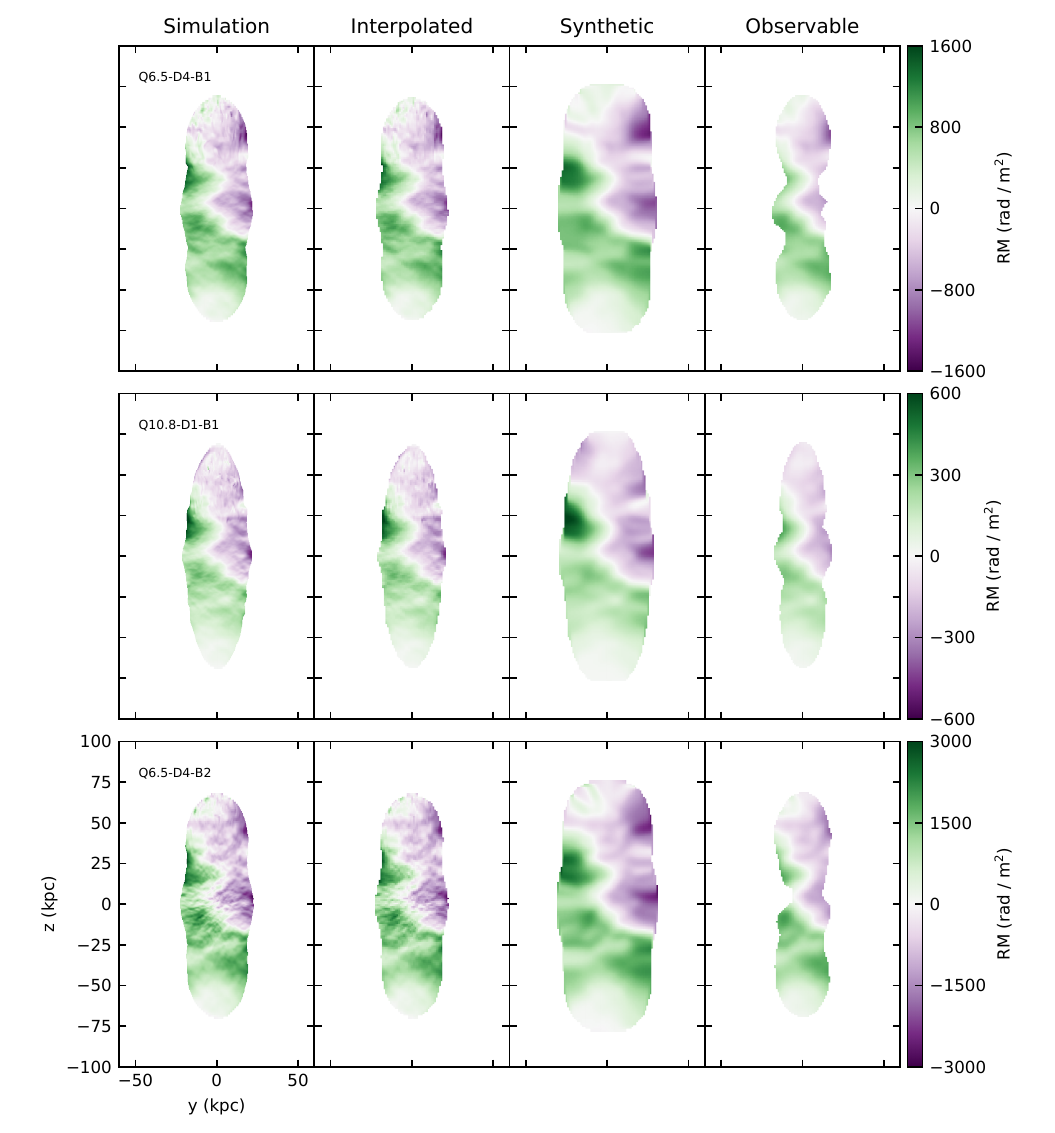}
    \caption{Comparison of RM maps for each synthetic observation as described in the text. From left to right: raw simulation data, this data interpolated to the instrument observing grid (resolution $1$ kpc/$0.889$ arcsec), this interpolated data convolved with $4.45$ arcsec beam, and in the final column this convolved data with a dynamic range of 10 at 151 MHz applied.}
    \label{fig:RM_observability}
\end{figure*}

We now consider which differences in the RM maps of our simulations are observable with a telescope. In Fig. \ref{fig:RM_observability}, we show the effects of sensitivity and resolution on our RM maps. On the far left, we show the raw simulation data, followed by this raw data interpolated to an observing grid with resolution $1$ kpc (corresponding to $0.889$ arcsec at $z = 0.056075$). In the third column we show this interpolated data convolved with a $4.45$ arcsec beam. The fourth column shows the visible data on the observing grid using a dynamic range of $10$ at $151$ MHz \footnote{Note that this dynamic range is for one source and does not account for other bright sources in a large field of view. This dynamic range should be consistent with lower surface brightness sources in a field with brighter sources.}; we consider this to be a lower limit on dynamic range to study here.

The interpolation to the instrument observing grid for resolutions of $1$ kpc or better does not impact the morphology of the RM map or the visibility of the FWHM in the RM distribution significantly. The primary difference between these synthetic observations and the raw simulation data is when the dynamic threshold is applied; if the instrument has low sensitivity, information mostly from the oldest electrons, primarily in the equatorial regions of the lobe, is lost. This will affect the most extreme RM values in the distribution, as shown in the top row of Fig. \ref{fig:structure_fns}, where the FWHMs of the RM distributions are largely unaffected by the differences in sensitivity. We estimate that most reasonable dynamic ranges at 151 MHz would be sufficient to capture the FWHM differences; we do not find significant departures from the distribution shape for dynamic ranges $\gtrsim 10$.

To be able to observe the RMs, we must also consider what frequencies the RMs will be visible at. This may be characterised using the depolarisation frequency, where the fractional polarisation drops to half of its high-frequency value, as given by \citet{krause_depolarisation_2007}:

\begin{equation}
\label{eqn:depolarisation}
    \nu_{1/2} = \sqrt{\frac{\sigma_{RM}}{13}} \: {\rm GHz},
\end{equation}

where $\sigma_{RM}$ is the RM dispersion, which we take to be the standard deviation in the RM integrated along each line of sight. Equation \ref{eqn:depolarisation} is based on an analysis of multiscale magnetic fields with high resolution data cubes as opposed to the classical \citet{burn_depolarization_1966} law, which assumes a two-scale field. 

\begin{figure*}
    \centering
    \includegraphics[width=\textwidth]{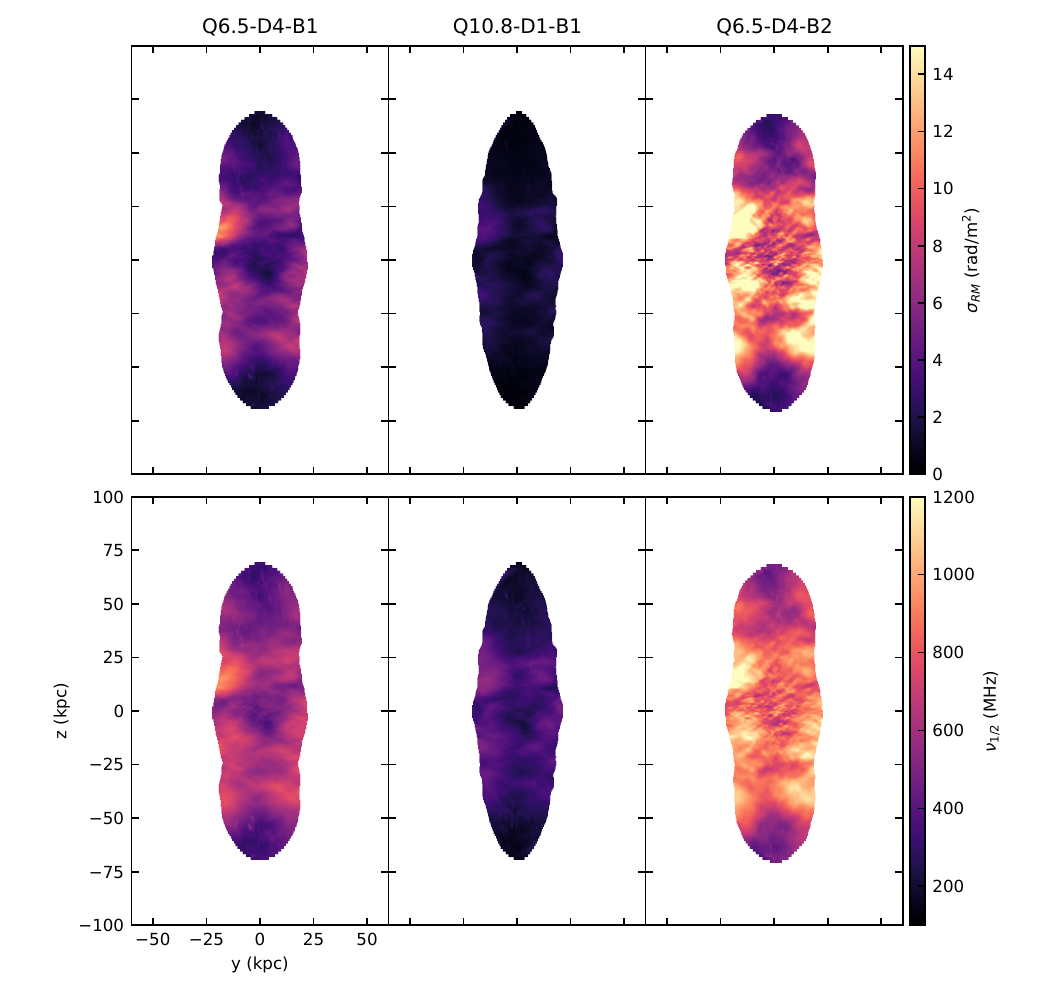}
    \caption{RM dispersions (top) and depolarisation frequencies (bottom) for simulations Q6.5-D4-B1, Q10.8-D1-B1, and Q6.5-D4-B2. Simulation Q6.5-D4-B1 is plotted at 15.0 Myr, simulation Q10.8-D1-B1 is plotted at 5.9 Myr, and simulation Q6.5-D4-B2 is plotted at 15.3 Myr.}
    \label{fig:depolar}
\end{figure*}

We show spatial RM dispersion and depolarisation frequency maps in Fig. \ref{fig:depolar}. The spatial structure of both $\sigma_{RM}$ and $\nu_{1/2}$ are similar to the total RM structures shown in Fig. \ref{fig:radio_sb_RM}, with enhanced regions to the sides of the lobes and a decrease towards the jet axis and lobe ends. This indicates that emission from the most recently shocked electrons near the jet hotspots will depolarise at lower frequencies than the electrons in the backflow. The depolarisation frequencies are $\lesssim 1$ GHz for each of our sources; we find the maximum and median depolarisation frequency in simulations Q6.5-D4-B1, Q10.8-D1-B1, and Q6.5-D4-B2 to be $955, 565, 1336$ and $517, 310, 839$ MHz respectively. Therefore, the RMs could only be entirely observed by instruments operating above $\sim 1.4$ GHz. However, since there are a range of different depolarisation frequencies across the source, the source will progressively depolarise, and the RM can still be characterised at MHz frequencies.

\begin{figure}
    \centering
    \includegraphics[width=\columnwidth]{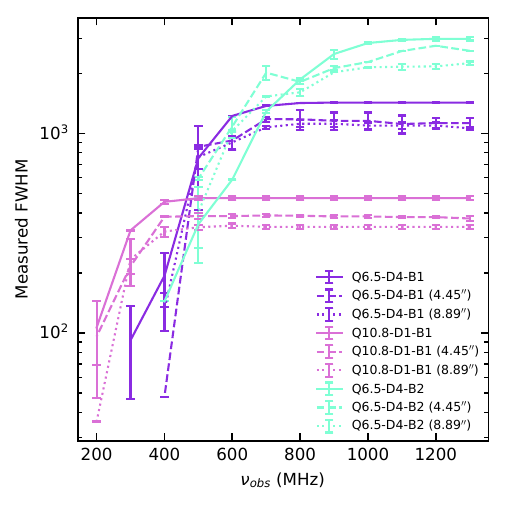}
    \caption{Measured RM FWHM vs observing frequency for simulations Q6.5-D4-B1, Q10.8-D1-B1, and Q6.5-D4-B2. Solid lines correspond to the raw simulation data with the depolarisation frequency threshold $\nu_{\rm{obs}} > v_{1/2}$ applied. Dashed $4.45^{\prime \prime}$ lines correspond to the `observable' RM maps that have been interpolated to a $1$ kpc observing grid, convolved with a $4.45$ arcsec beam, with a dynamic range threshold of 10 and the depolarisation frequency threshold applied. Dotted $8.89^{\prime \prime}$ lines correspond to the same `observable' RM maps, but convolved with a $8.89$ arcsec beam instead.}
    \label{fig:measured_FWHM}
\end{figure}

We quantify the observability of the FWHM of the RM distribution at MHz frequencies in Fig. \ref{fig:measured_FWHM}. We plot the measured FWHMs of the RM distributions where $\nu_{\rm obs} > v_{1/2}$ for the raw simulation RM and `observable' RM maps shown in Fig. \ref{fig:RM_observability}. We also include an `observable' RM map convolved with a $8.89$ arcsec beam. To estimate the error in our FWHM measurement, we calculate the FWHM with four tolerance levels corresponding to $0.04, 0.06, 0.08$ and $0.1$ times the distribution peak, taking the mean and standard deviation of these measurements. For all three simulations, we do not find any observable polarised emission at frequencies $\lesssim 200$ MHz. FWHM values converge for all sources as frequency increases, and the source RM becomes fully visible. Each of our sources begin to lose RM information at lower frequencies, so the measured FWHM falls away from the true value. The frequencies at which the measured FWHM are more than one standard deviation away from the true value for our three simulations are $\sim 700, 300,$ and $1000$ MHz for simulations Q6.5-D4-B1, Q10.8-D1-B1, and Q6.5-D4-B2 respectively. We note that powerful FR-IIs in poor environments are likely to be more reliably observed at lower frequencies, and stronger cluster magnetic fields result in stronger depolarisation at higher frequencies. These different depolarisation signatures can provide key information about the environment the source resides in.

For these sources observed at 151 MHz, the lower-powered jets will not be polarised. The higher-powered jet is partially polarised at this frequency, but the measured FWHM is well below the true value. To distinguish between these sources at 151 MHz we would require polarimetry at higher frequencies. For frequencies $\gtrsim 1$ GHz, synchrotron losses will distinguish these particular sources in their surface brightness, but in general, the size-luminosity degeneracy may occur at higher frequencies for different jet powers and environments. The pattern of the measured FWHM of the RM distribution with observing frequency still encodes much information about the environments, even without reconstructing the full RM distribution.

Detecting the differences between different jet power and environment density combinations, as we have shown in this paper, requires sufficient resolution to detect the RM distribution. The $4.45$ arcsec resolution used in this paper is optimistic for large surveys of AGN jets (Table \ref{tab:polsurveys}). We see in Fig. \ref{fig:measured_FWHM} that a reduction in observing resolution results in a smaller measured FWHM. However, there are clear differences between the environments considered in our simulations that may still be detected at a lower resolution, if the RM resolution is sufficient. Table \ref{tab:polsurveys} lists the key parameters for various polarisation surveys, including the RM resolution (i.e. the RMSF FWHM), which ranges from 1.2 rad/m$^2$ for LoTSS to $370$ rad/m$^2$ for Apertif, and the maximum RM scale, which ranges from $0.97$ rad/m$^2$ for LoTSS to $2500$ rad/m$^2$ for QUOCKA. For all listed surveys other than LoTSS and POGS, the RM dispersions shown in Fig. \ref{fig:depolar} are lower than this maximum scale. Hence, for most current surveys, the RM resolution and maximum scale are sufficient to detect the differences between our simulated sources.

With these current surveys, the detection of RMs is primarily limited by survey resolution. However, for poorly resolved sources, \citet{anderson_study_2016} show that broadband depolarisation modelling can discern information about the RM distribution at scales beneath the observing beam. This would allow for the detection of the FWHMs and the differences between them in current surveys, particularly for sources at high redshift. Upcoming sky surveys using next-generation instruments such as SKA-low and the ngVLA will have sufficient resolution to study large populations of AGN jets in detail and characterise their environments using polarimetry. In particular, the SKA-low survey will have an RM resolution of $\sim 0.1$ rad/m$^2$ \citep{2008A&A...480...45S} and so will present an opportunity to study even smaller differences in RM than our current surveys.

\begin{table*}
	\centering
	\caption{Key parameters for various polarisation surveys in progress. References: 1) Gaensler et al. \emph{in prep.}, 2) \citet{2022A&A...663A.103A}, 3) \citet{2022A&A...659A...1S}, 4) \citet{2020PASA...37...29R}, 5) Heald et al. \emph{in prep.}, 6) \citet{2020PASP..132c5001L}.}
	\label{tab:polsurveys}
	\begin{tabular}{lcccccccc}
		\hline
		\hline
		Survey & Frequency & Coverage & 1-$\sigma$ rms & Resolution & RMSF FWHM & Max. RM & Max. scale & Reference \\
		& (MHz) && ($\mu$Jy beam$^{-1}$) & (arcsec) & (rad m$^{2}$) & (rad m$^{2}$) & (rad m$^{2}$) & \\
		\hline
        POSSUM  & 800$-$1088  & 2pi sr & 18 & 13 & 54 &4000&40& 1\\
        POSSUM band 2  & 1130$-$1430 & 1.5pi sr &30&11&370&4000&72& 1\\
        Apertif & 1130$-$1430 & Irregular &16&15&370&4000&72& 2\\
        LoTSS   & 120$-$168   & -1 $<\delta<$ +90&70&6&1.2&450&0.97& 3\\
        POGS    & 216         & -82 $<\delta<$ +30& Varies; $1\times10^3$--$3\times10^4$ & 180 &6.2&1000&1.9&4 \\
        QUOCKA  & 1300$-$8500     & 
        Targeted pointings&50&Flexible; 1--15&67&90,000&2500& 5\\
        VLASS   & 2000$-$4000     & -40 $<\delta<$ +90&69&~5& 200&2300&560& 6\\
		\hline
	\end{tabular}
\end{table*}

\subsubsection{Structure functions}

\begin{figure*}
    \centering
    \includegraphics[width=\textwidth]{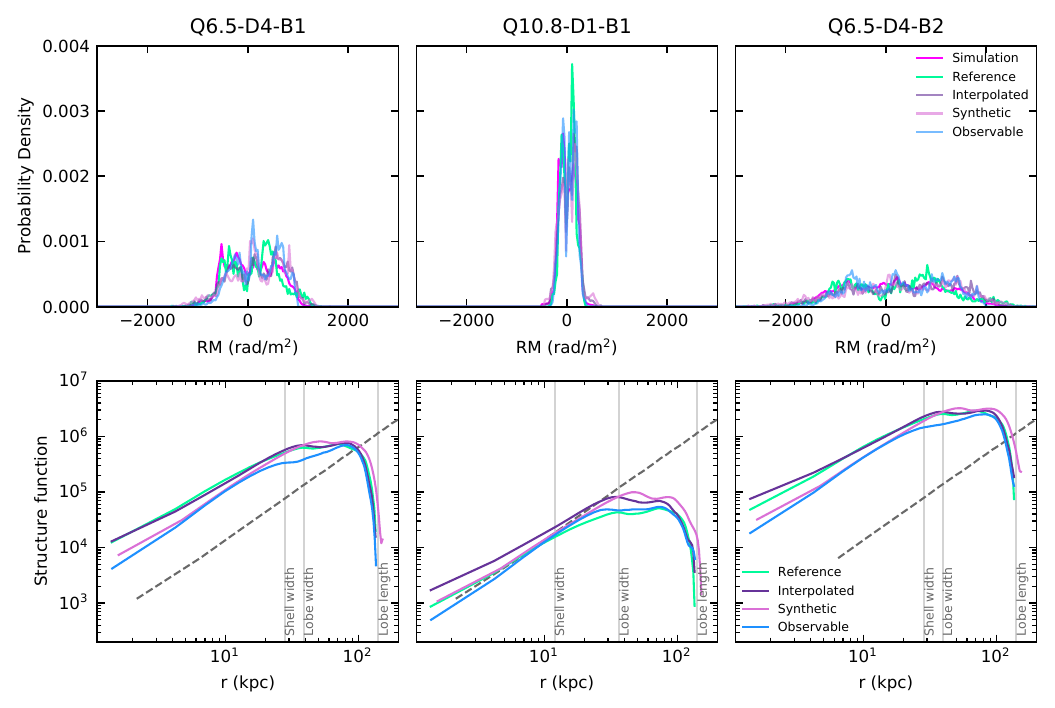}
    \caption{Histograms (top) and structure functions (bottom) for the different synthetic observations as shown in Fig. \ref{fig:RM_observability}, as well as the unaffected environment RM shown in Fig. \ref{fig:RM_components}. The `reference' RM is interpolated to the same observing grid as the `interpolated' RM (resolution $1$ kpc/$0.889$ arcsec) for the purposes of plotting its structure function. The `synthetic' RM is the `interpolated' RM convolved with a $4.45$ arcsec beam and the `observable' RM corresponds to this convolved data with a dynamic range of 10 at 151 MHz applied, as seen in Fig. \ref{fig:RM_observability}.}
    \label{fig:structure_fns}
\end{figure*}

If the spatial RM distribution of an observed source can be measured, we can determine its structure function. The structure function describes the two-dimensional fluctuations of the RM and is given by \citep{laing_structures_2008,guidetti_magnetized_2012,minter_observation_1996,simonetti_small-scale_1984}: 

\begin{equation}
    S(r) = \langle [RM(r + r') - RM(r')]^2 \rangle,
\end{equation}

where $r$ and $r'$ are vectors in the plane of the sky and $\langle \rangle$ is an average over $r'$. In the bottom row of Fig. \ref{fig:structure_fns} we show the structure functions for the `interpolated', `synthetic', `observable' and `reference' RMs shown in Figs \ref{fig:RM_components} and \ref{fig:RM_observability}. The `interpolated' RM corresponds to the raw simulation RM interpolated to the same $1$ kpc observing grid as the synthetic RM, but it has not been convolved with a beam. Likewise, the `reference' RM has been interpolated to the same observing grid. We compare our structure functions to the Kolmogorov spectrum structure function as given by \citep{laing_structures_2008}:

\begin{equation}
    S(r) = \frac{4 \pi^{q-1}}{q - 2} \frac{\Gamma(2 - q/2)}{\Gamma(q/2)} r^{q-2},
\end{equation}

where $\Gamma$ is the Gamma function, and $2.5 < q < 4$ with $q = 11/3$ for a Kolmogorov spectrum.

The structure functions all roughly follow a power-law increasing slope until the lobe-width scale (second vertical grey line from the right), after which the structure function begins to turn over. The `observable' RM has some differences compared to the `interpolated' and `synthetic' RMs. The power in the `observable' structure function tends to decrease at scales just below the lobe width, with the full signal in the `interpolated' and `synthetic' RMs remaining a power law until the lobe width is reached. This initial power law does not follow the Kolmogorov spectrum (grey dashed lines) in most cases; this is likely due to the simulation resolution and interpolation of the RM.

The structure functions all rapidly decrease towards the lobe-length scale (third vertical grey line from the right), as the number of $(r,r')$ pairs to average over decreases. This flattening from power-law structure has been observed in observations \citep{laing_structures_2008,guidetti_ordered_2011,guidetti_magnetized_2012}, however, unlike \citet{2011MNRAS.418.1621H} we find that this change in gradient is not the effect of source expansion. For the lower-powered jets, we show in Fig. \ref{fig:structure_fns} that the `interpolated' and `synthetic' RMs do not depart significantly from the `reference' RM structure function. Instead, we find it is the shape of the jet cocoon that influences the RM structure function. This function only changes significantly for the `observable' case, where the RM map is dynamic-range limited and loses the full jet cocoon shape (see Fig. \ref{fig:RM_observability}). We also note that the `synthetic' RM structure function has its initial turnover at a slightly larger scale than the `interpolated' and `reference' RMs due to the beam convolution smoothing out the RM map to larger scales as seen in Fig. \ref{fig:RM_observability}.

For the higher-powered jet in simulation Q10.8-D1-B1, the `interpolated' and `synthetic' structure functions are amplified above the `reference' structure function. This indicates that the higher-powered jet can amplify the RM values above the `reference' level more significantly than the lower-powered jets. We conclude that this is due to the greater compression and amplification in the Faraday screen (as discussed in Section \ref{section:dynamics}), in agreement with the findings of \citet{2011MNRAS.418.1621H}. However, for the dynamic range-limited `observable' structure function, this amplification above the `reference' level is lost. This highlights the importance of sensitivity in detecting the true underlying structure function associated with a source.

The left-most grey vertical line in each bottom panel of Fig. \ref{fig:structure_fns} corresponds to the length scale of the width of the Faraday screen (at $z=0$). For the lower-powered jets, this length scale seems to be associated with the spectrum departing from the power-law. This length scale does not seem to be associated with any particular feature of the structure functions in simulation Q10.8-D1-B1.

In general, the departure from power-law structure does not always result in a turn-down in the spectrum and depends on the structures of the cluster magnetic field and density profiles \citep[e.g.][]{2011MNRAS.418.1621H}. However, since these profiles are the same in all simulations, the overall shape of the structure function is very similar (accounting for differences in jet shape). The decrease in central density can be seen in the decreased magnitude of the structure function for simulation Q10.8-D1-B1; likewise, the increased magnetic field can be seen for Q6.5-D4-B2.

In simulation Q6.5-D4-B2, the `interpolated' structure functions at the small-scale end are amplified above the `reference' structure function. This is likely due to the central high-resolution grid patch influencing the evolution of the magnetic field, which occurs at higher ambient magnetic field strengths. This results in rapidly changing RM magnitudes on small scales, primarily along $y = 0$ and $z = 0$ (see Figs \ref{fig:RM_components} and \ref{fig:RM_observability}). These fluctuations occur on the smallest scale of the simulation grid, thus boosting the structure function at this end of the spectrum, even when interpolated to the coarser observing grid. We expect this effect to be purely numerical and not physically motivated.

\section{Discussion}
\label{section:discussion}

\subsection{Improvements and limitations of our methods}
\label{section:improvements_limitations}

The methods presented in our paper extend previous work. We confirm the results of \citet{2011MNRAS.418.1621H} with higher resolution simulations. Using PRAiSE \citep{yates-jones_praise_2022}, we have made surface brightness images including full synchrotron, adiabatic, and inverse-Compton losses. We extend the RM analysis of \citet{2011MNRAS.417..382H,2011MNRAS.418.1621H} and use a dynamic range threshold for these surface brightness images to create `observable' spatial RM maps. We have considered the depolarisation of our sources and the feasibility of observing these sources and their RMs to study AGN environments.

There are some drawbacks to our methods in this paper, primarily related to the modelling of the cluster magnetic field. The stretched grid in our simulations models the jets from pc to kpc scales well. However, the cluster magnetic field is generated on a uniform grid and then interpolated to this stretched grid. The non-uniform grid cells on the simulation grid influence the evolution of the magnetic field; this occurs on small scales in the high-resolution grid patches along the axes. This effect is most visible in the RM map for simulation Q6.5-D4-B2 (Fig. \ref{fig:radio_sb_RM}). We also found that our simulated cluster magnetic fields resulted in non-isotropic RMs (Section \ref{section:RM_probe_env}). Finally, we note that the integration along the RM line of sight may result in numerical errors due to the nature of integrating a grid along an axis at some angle to the coordinate axes. We stress the importance of modelling the cluster magnetic field well and highlight this as an area for future improvements.

We also note that we have used a shock threshold corresponding to a Mach number $M \sim 1.18$. This is in contrast with the $M \sim 2.24$ threshold that is used in \citet{yates-jones_praise_2022}, which corresponds to a critical Mach number above which particle acceleration is most likely to occur \citep{2014ApJ...780..125V,2019ApJ...876...79K,2022ApJ...925...88H}. This shock threshold changes the final luminosity of the source, which will in turn affect the size-luminosity tracks. We found that for the higher shock threshold, the size-luminosity track of the source in simulation Q10.8-D1-B1 was virtually unchanged; however, for the lower-powered jets, the size-luminosity tracks were reduced by $\sim 20 \%$ for projected source lengths above $\sim 100$ kpc. This reduction in luminosity is primarily due to a weaker shock in the hotspot region of these lower-powered jets, resulting in less emission at the jet head. By using the lower shock threshold, the sources in simulations Q6.5-D4-B1 and Q6.5-D4-B2 retain hotspot-like structures and therefore more closely resemble FR-II sources. The details of the particle acceleration are not critical to the results of this paper.

\subsection{Implications for AGN jet feedback}
\label{section:feedback}

The demonstrated ability of RM analysis to distinguish between different combinations of jet power and environments has significant implications for the quantification of AGN jet feedback. This feedback is done in two ways for powerful FR-II jets: their bow shock heats the intracluster medium (ICM) and also lifts this gas out of the gravitational potential well \citep[e.g.][]{2002MNRAS.335..610A,2013MNRAS.430..174H,2014MNRAS.443.1482H}. However, the effects of these mechanisms are impacted by the relationship between jet power and environment.

Despite very similar observable radio continuum features, the different jets in our simulations provide different amounts of feedback to the surrounding gas. As Table \ref{tab:amplification_ratios} shows, the jets in the higher density environment (simulations Q6.5-D4-B1 and Q6.5-D4-B2) sweep up and shock heat a larger mass of the surrounding gas: 20\% of the gas within 150 kpc of the host galaxy for the denser environment, but only 9\% for the less dense environment. We cannot determine the difference in the amount of gas affected by feedback without information on the type of environment the source resides in.

Jets in less dense environments (simulation Q10.8-D1-B1) are more powerful for the same radio luminosity, and the resultant lobes expand faster than in the low-power/high-density case (simulations Q6.5-D4-B1 and Q6.5-D4-B2). The high-power jet heats the shocked gas to higher temperatures (an increase of 110\%, compared to an 18\% increase for the lower-power simulations). However, the shocked gas in simulation Q10.8-D1-B1 still has a lower density than the low-power jet simulations (Table \ref{tab:amplification_ratios}). The combination of higher temperature and lower density results in much longer cooling times for the high-power jets, meaning more efficient negative feedback - albeit delivered to smaller masses of gas. This theoretical finding is consistent with the observational result that FR-II sources in groups provide longer-lasting feedback than cluster sources \citep{2011MNRAS.413.2815S}.

These differences in the temperature and amount of gas affected by jet feedback cannot be distinguished using radio luminosity and size alone. As RM distributions can successfully distinguish between different environments, this approach may lead to more accurate estimates of AGN jet feedback.

\section{Conclusions}
\label{section:conclusions}

In this paper, we presented three-dimensional relativistic magnetohydrodynamic simulations of Cygnus A-like powerful radio galaxies in turbulent external magnetic fields. Using three different combinations of jet and environment parameters, we present a solution to the degeneracy in the relationship between the size and luminosity of radio galaxies. By using Faraday rotation measures, we find that a higher environment density or a higher cluster magnetic field strength broadens the distribution of RM values in a given source, allowing jet environments to be characterised using the RM for the given density and magnetic field distributions. We also confirmed the findings of \citet{2011MNRAS.418.1621H} with higher resolution simulations. We found that the jets act to align an underlying RM reversal with the jet axis, a reversal that is also seen in observations \citep{2020ApJ...903...36S,anderson_spiderweb_2022}. Our results stress the importance of modelling the external magnetic field accurately. 

The effects of the environment on the FWHM of the RM distribution are visible in current and future surveys if the survey resolution is sufficient. We found that observing grid resolution better than $1$ kpc ($=0.889$ arcsec at redshift $z=0.056075$) is sufficient to model the RM distribution for our simulated sources. We apply a dynamic range threshold of $10$ at 151 MHz to our RM maps. This threshold preferentially excludes information from the lobe edges (primarily in the equatorial region), where the jet enhances the RM the most. These thresholds used do not greatly affect the shape of the RM distribution, so any reasonable dynamic range should be sufficient to study similar Cygnus A-like sources. Current surveys such as POSSUM \citep{2010AAS...21547013G} and VLASS \citep{2020PASP..132c5001L} have sufficient RM resolution to sample RM distributions for sources similar to our simulated ones. For sources with insufficient surface brightness resolution, such as those at higher redshifts than our simulated sources, broadband polarimetry has the potential to model the RM distributions \citep{anderson_study_2016}. Future surveys with SKA-low \citep{2008A&A...480...45S} and the ngVLA \citep{2023AAS...24135702S} will have the capability to characterise AGN jet environments in more detail than current surveys.

The full RM maps discussed here are only visible for an observing frequency $\nu_{\rm obs} \gtrsim 1$ GHz for our simulated sources. The maximum depolarisation frequency ranges between $565$ and $1136$ MHz for our simulated sources. We found that the depolarisation frequency decreases towards the jet axis and hotspots, indicating that the emission from recently shocked electrons in these regions will depolarise at a lower frequency. We measure the FWHM of the observable RM distribution for MHz frequencies, finding that the measurement becomes reliable at frequencies between $300 - 1000$ MHz for our three simulations. We note that a stronger cluster magnetic field strength results in depolarisation at higher frequencies.

The structure functions of our RM maps follow the Kolmogorov power-law slope for scales smaller than the lobe width. After this point, the spectrum turns over, decreasing sharply at the lobe length scale. We found that our dynamic range threshold causes non-power-law behaviour at smaller length scales than the lobe width due to the removal of amplified RM values at the edges of the jet lobe. We also found that small numerical effects influenced the structure function of the high magnetic field simulation; this again stresses the importance of modelling the magnetic field well. Our simulations show that changes in the structure function of the lower-powered jets can be primarily explained by the environment and shape of the jet cocoon, rather than the effects of source expansion on the environment. In contrast, the higher-powered jet amplifies the RM structure function above the `reference' ($t = 0$ Myr) level. This amplification is not seen when the dynamic range threshold is applied.

The differences between our simulations have implications for AGN feedback. We find that our high-powered jet in a low-density environment, with a narrow rotation measure distribution, affects a lower percentage of environment gas, but it is more effective at increasing the temperature of and therefore changing the thermal state of the gas. Thus, combining Stokes I and Faraday rotation observations of radio galaxies opens a new window to studying AGN feedback in large surveys.

In this work, we have focused on using the RMs as a probe of environment. In future work, we will present synthetic observations using full Stokes' parameters to study the polarisation properties of AGN jets. We will modify the synthetic emission code PRAiSE \citep{yates-jones_praise_2022} to use the magnetic field in the jet cocoon to model our radiative losses and study the differences between the radio continuum properties of RMHD and RHD jets. This and future work will be used in the CosmoDRAGoN project \citep{2023PASA...40...14Y} to study the synthetic emission of RMHD jets in realistic, magnetized, cluster environments.

\section*{Acknowledgements}

LJ thanks the University of Tasmania for an Australian Government Research Training Program (RTP) Scholarship. PR thanks the University of Tasmania for the Don Gaffney Scholarship and the Gates Cambridge Trust for research funding. This research was carried out using the high-performance computing clusters provided by Digital Research Services, IT Services at the University of Tasmania. We thank \citet{2022NatAs...6.1021K} for their guide to writing astronomy papers. We acknowledge the support of the developers providing the Python packages used in this paper: Astropy \citep{2022ApJ...935..167A}, JupyterLab \citep{2016ppap.book...87K}, Matplotlib \citep{2007CSE.....9...90H}, NumPy \citep{2020Natur.585..357H}, and SciPy \citep{2020SciPy-NMeth}.

\section*{Data Availability}

The data underlying this article will be shared on reasonable request to the corresponding author.



\bibliographystyle{mnras}
\typeout{}
\bibliography{references} 




\bsp	
\label{lastpage}
\end{document}